\documentstyle[preprint,prl,aps,floats,epsf]{revtex}

\tightenlines

\def\simlt{\stackrel{<}{{}_\sim}}

\newcommand{\md}{m_\Delta}
\newcommand{\ms}{m_*}
\newcommand{\be}{\begin{eqnarray}}
\newcommand{\ee}{\end{eqnarray}}
\newcommand{\ben}{\begin{eqnarray*}}
\newcommand{\een}{\end{eqnarray*}}

\begin{document}

\draft
\title{The Role of Nucleons in Electromagnetic Emission Rates}

\author{{\bf James V. Steele}$^1$ and {\bf Ismail Zahed}$^2$} 

\address{$^1$Department of Physics, The Ohio State University,
Columbus, OH 43210, USA\\
$^2$Department of Physics and Astronomy,
SUNY, Stony Brook, NY 11794, USA}
\date{February 8, 1999}
\maketitle

\begin{abstract}
Electromagnetic emission rates from a thermalized hadronic gas are
important for the interpretation of dilepton signals from heavy-ion
collisions. 
Although there is a consensus in the literature about rates for a pure
meson gas, qualitative differences appear with a finite baryon
density.
We show this to be essentially due to the way in which
the $\pi N$ background is treated in regards to
the nucleon resonances.
Using a background constrained by unitarity and broken chiral symmetry, it is
emphasized that the thermalized hadronic gas can be considered dilute.
\end{abstract}

\thispagestyle{empty}

\newpage

Recent relativistic heavy-ion collisions at CERN have reported an
excess of dileptons over a broad range of invariant mass 
$M=300$-$500$~MeV \cite{CERES,HELIOS}. 
Various theoretical assessments of the dilepton 
emission rates have tried to understand this excess. 
In the absence of baryons, the 
rates achieved using detailed reaction processes~\cite{GALE,ALLR}
and spectral sum rules~\cite{US1,ALLS} agree with each other, but fail
to reproduce the enhancement in the data. In the 
presence of baryons, calculations using 
many-body dynamics~\cite{RAPP,RAPP1} lead to
larger rates below the $\rho$ peak compared to results from 
spectral considerations constrained by broken 
chiral symmetry~\cite{US2}.  

\begin{figure}
\begin{center}
\leavevmode
\hbox{
\epsfxsize=3.25in
\epsffile{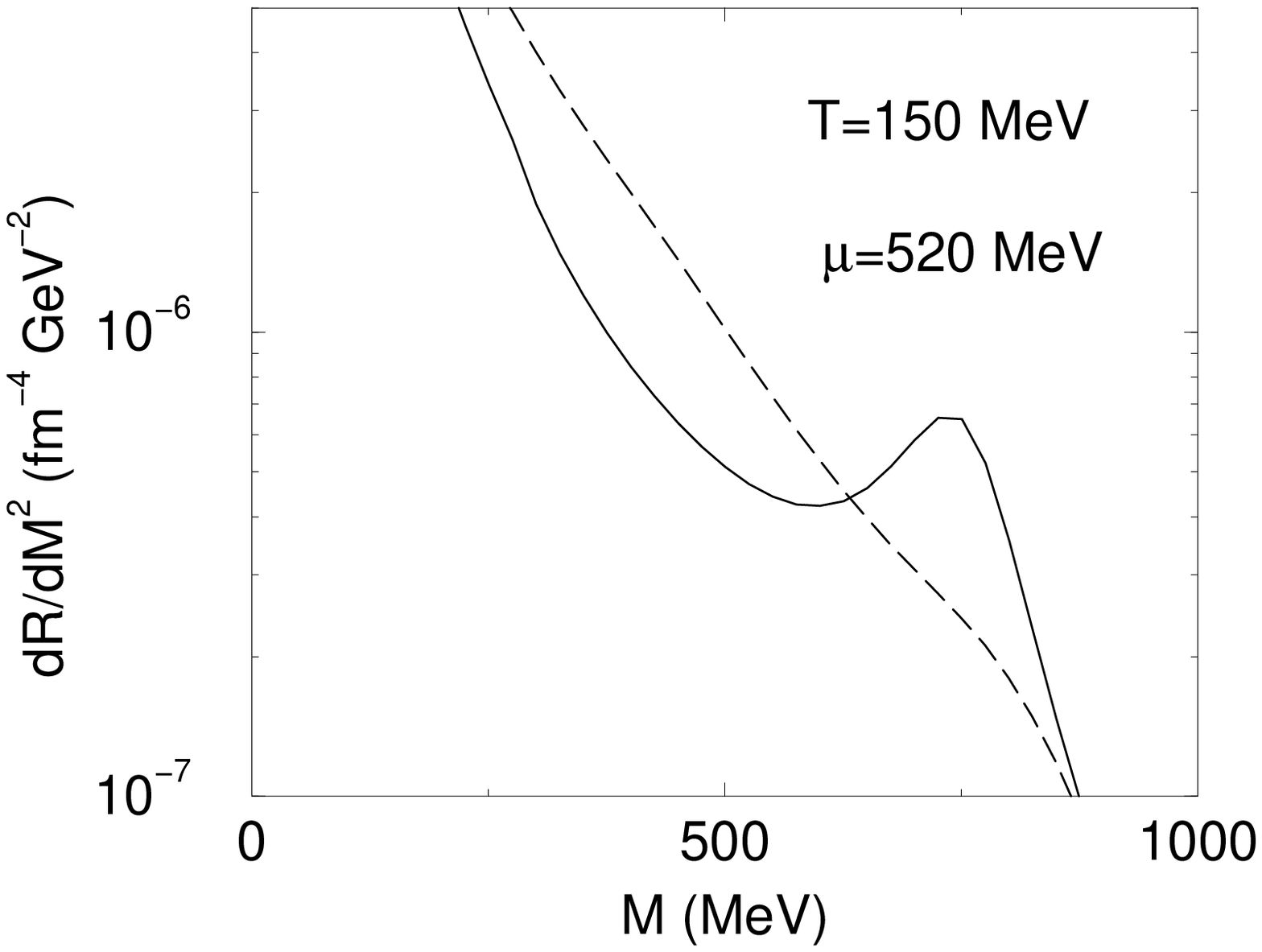}
\epsfxsize=3.25in
\epsffile{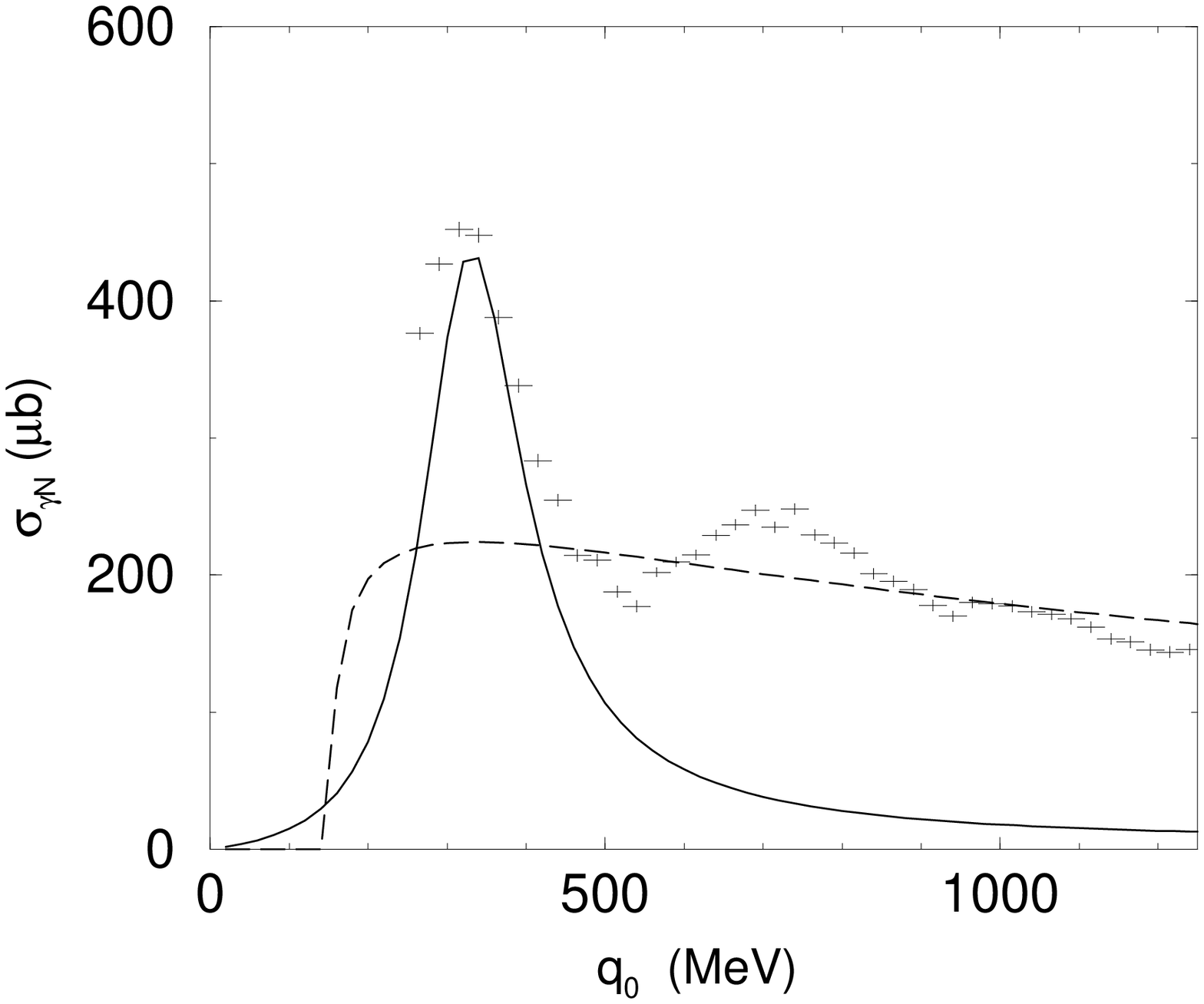}
}
\end{center}
\caption{\label{rappcomp} 
The left plot compares nucleon rates from Refs.~\protect\cite{US2}
(solid) and \protect\cite{RAPP} (dashed).
The right plot shows the results for the (isospin averaged) 
Compton scattering on the nucleon from the $\Delta$ (solid) and
background (dashed) versus data~\protect\cite{PDG}.}
\end{figure}

This is
illustrated by the left plot of Fig.~\ref{rappcomp}
for pertinent temperature $T=150$ MeV and
baryon chemical potential\footnote{
We express our rates using a complete set of stable states
\cite{US2} and 
therefore $\mu$ refers solely to nucleons. 
In the analysis of Ref.~\cite{RAPP,RAPP1},
$\mu$ refers to all possible baryons as they retain the unstable 
particles in their final states. The rate comparison done here
is only meaningful for fixed $T$ and $\mu$. 
\vspace*{.4cm} %added for nice spacing on the following pages
}
$\mu=520$~MeV. 
Although
neither rate accommodates the data in a realistic
hydrodynamical evolution~\cite{PRA},  the rates 
from Ref.~\cite{RAPP,RAPP1} are about two to three times 
larger than those of Ref.~\cite{US2}, being 
within only two standard deviations of the data.
We show below that the discrepancy between these two rates
originates chiefly from the $\pi N$ background, and that the hadronic 
gas is essentially dilute, in confirmation of our earlier
work~\cite{US2}. 

Although perturbative unitarity fixes the $\pi N$ background uniquely
in the Compton amplitude, its extrapolation away from the photon-point
requires care. In Ref.~\cite{US2} this extrapolation was implemented by paying
due care to threshold unitarity. In the future, the complicated issue of
background versus resonances should be properly resolved 
by enlarging the theoretical analysis
to pion photo-production as well as pion knock-out
while obeying unitarity and broken chiral symmetry as
detailed in Ref.~\cite{MASTER}.
For now, we concentrate on illuminating the reason for the above differences
in the two theoretical estimates containing nucleons.

In a thermal equilibrated hadronic gas, the rate ${\bf 
R}$ of dileptons produced in an unit four volume follows from the thermal 
expectation value of the electromagnetic current-current correlation 
function \cite{LARRY}. For massless leptons with momenta $p_1$ and $p_2$,
the rate per unit invariant momentum $q =p_1+p_2$ is given by \cite{US2}
\be
\frac {d{\bf R}}{d^4q} = -\frac{\alpha^2}{3\pi^3 q^2}\,
\,\frac 1{1+e^{q^0/T}}\,\,{\rm Im}{\bf W}^F (q)
\label{1}
\ee
where $\alpha =e^2/4\pi$ is the fine structure constant, and
\be
{\bf W}^F (q) = i\! \int\!\! d^4x \, e^{iq\cdot x} \,
{\rm Tr} \left(e^{-({\bf H}-\mu\, {\bf N} -\Omega)/T} \,\,T^*{\bf
J}^{\mu} (x){\bf J}_{\mu} (0)\right)\, .
\label{4}
\ee
$e{\bf J}_{\mu}$ is the hadronic part of the electromagnetic current,
${\bf H}$ is the hadronic Hamiltonian, $\mu$ is the baryon chemical
potential, $\bf N$ is the baryon number operator, $\Omega$ is the Gibbs
energy, $T$ is the temperature, and the trace is over a complete set of
hadron states.

We expand the trace in Eq.~(\ref{4}) using pion and nucleon
states.  This is justified for temperatures $T\simlt m_{\pi}$ 
and final nucleon densities $\rho_N\simlt 3\rho_0$ with
$\rho_0=0.16$~fm$^{-3}$, since the relevant
expansion parameter $\kappa$ for each particle is less than $\frac13$
in that regime.
This allows us to only retain terms up to first order in density
\cite{US2}
\be
{\rm Im} \,{\bf W}^F (q) = -3q^2 {\rm Im} \;{\bf \Pi}_V(q^2)
+\frac1{f_\pi^2}\int\! d\pi\, {\bf W}^F_\pi(q,k) 
+ \int\! dN\, {\bf W}^F_N (q,p) 
+ {\cal O}\!\left(\kappa_\pi^2, \kappa_N^2, \kappa_\pi\kappa_N
\right). 
\label{6}
\ee
The phase space factors are 
\be
dN = \frac {d^3p}{(2\pi)^3} \frac 1{2E_p}
\frac 1{e^{(E_p-\mu)/T} +1}\qquad{\rm and}\qquad
d\pi = \frac {d^3k}{(2\pi)^3} \frac {1}{2\omega_k} 
\frac 1{e^{\omega_k/T} -1}\, ,\nonumber
\ee
with nucleon energy $E_p=\sqrt{m^2 +p^2}$ and pion energy 
$\omega_k=\sqrt{m_{\pi}^2 +k^2}$.  The first term in Eq.~(\ref{6}) is the
transverse part of the isovector correlator $\langle 0|T^* {\bf V}{\bf V} |0
\rangle$ and summarizes the results of the resonance gas model. It is
given by $e^{+}e^{-}$ annihilation data. At low and moderate $q^2$ 
this is dominated by the $\rho$ and $\rho'$ while at high $q^2$
its tail is determined by the $q\overline{q}$ spectrum.

The term linear in pion density can be expressed in terms of
experimentally measurable quantities by use of chiral
reduction formulas.
The important contributions are \cite{US1}
\be
{\bf W}^F_{\pi}(q,k) \simeq&& 12q^2 {\rm Im}\; {\bf \Pi}_V(q^2)
-6(k+q)^2 {\rm Im}\; {\bf \Pi}_A\left( (k+q)^2 \right) +
(q\rightarrow -q)
\nonumber\\
&&{}+8((k\cdot q)^2-m_\pi^2 q^2)\; {\rm Im}\;
{\bf \Pi}_V(q^2) 
\;{\rm Re} \left( \Delta_R(k+q) + \Delta_R(k-q) \right)  
\label{7}
\ee
with $\Delta_R (k)$ the retarded pion propagator
and ${\bf \Pi}_A$ the transverse part of the iso-axial correlator $\langle
0 | T^* {\bf j}_A {\bf j}_A| 0\rangle$ which follows from tau decay
data \cite{US1}.  It is dominated by the $a_1$ resonance.

The term linear in nucleon density is
just the spin-averaged forward Compton scattering amplitude on the
nucleon with virtual photons.  This is only measured for various
values of $q^2 \le 0$.  However, the dilepton and photon rates require
$q^2 \ge 0$.  Therefore, only the photon rate for this term can be
determined directly from data by use of the optical theorem
\be
e^2 {\bf W}^F_N(q,p) =
-4(s-m^2) \sum_I \sigma_{\gamma N}(s)
\label{8}
\ee 
with $s=(p+q)^2$. 
For off-shell photons, we must resort to chiral constraints to
determine the nucleon contribution to the dilepton rate.  Broken chiral
symmetry dictates uniquely the form of the strong interaction
Lagrangian (at tree level) for spin-$\frac12$ particles. 
Perturbative unitarity follows
from an on-shell loop-expansion in $1/f_{\pi}$ that enforces 
current conservation and crossing symmetry. To one-loop, the $\pi N$
contribution is parameter free. The large contribution of the
$\Delta$ to the Compton amplitude near threshold  is readily 
taken into account by adding it as a unitarized tree term
to the one-loop result \cite{US2}. The result for Compton scattering
on the nucleon from Ref.~\cite{US2} is shown in the
right plot of Fig.~\ref{rappcomp} versus data~\cite{PDG}.

This fit to the Compton data is good, given the fact it is a
parameter-free analysis.  However, we must 
determine the role of the $N^*(1520)$, since about 20\% 
of the cross section is unaccounted for in that kinematic region.
More importantly, it has been suggested that
the decay of this resonance in matter
is what feeds the dilepton rate enhancement~\cite{RAPP,RAPP1}.
We note that the $N^*(1520)$ has 
about a 50\% branching ratio to the $\pi N$ channel, making it 
difficult to disentangle
from the $\pi N$ background.

The contribution of the $N^*(1520)$ to the Compton amplitude
follows readily from the transition matrix element
\be
\langle N(p) | {\bf J}_\mu | N^*(k) \rangle &=& \overline{u}(p)
\Bigg[ Q_*(q^2) (\gamma_\mu q_\nu - g_{\mu\nu} \rlap/q)
+iS_*(q^2) \sigma_\mu^\lambda q_\lambda q_\nu 
\nonumber\\
&&{}\qquad
+\left( R_*(q^2) + \rlap/q \overline{R}_*(q^2) \right) (q_\mu q_\nu -
g_{\mu\nu} q^2 )
\Bigg] \frac{1+\tau^3}2
u^\nu (k) \, .
\label{NS1}
\ee
The form factors $Q_*, S_*, R_*$ and $\overline{R}_*$ comprise a maximal
set and are real by time reversal invariance. 
This decomposition parallels the one adopted
for the $\Delta$ in Ref.~\cite{US2} with adjustments for the difference in
parity and isospin between the two resonances. 
Hence, the $N^*(1520)$ contribution to
the Compton amplitude ${\cal M}_*(s,q^2)$ is the same form as
${\cal M} (s,q^2)$ for the $\Delta$ quoted in Ref.~\cite{US2} as long as
$\md$ is changed to $-\ms$ (parity) and the prefactor
$\frac 43$ is changed to $1$
(isospin). 
Just as for the $\Delta$, the form factors in Eq.~(\ref{NS1}) 
will be assumed $q^2$ independent. 

The resulting couplings are constrained
by the nucleon polarizabilities, the $E/M$-ratio, and the Compton
amplitude.  The nucleon polarizabilities can be separated into
the loop, $\Delta$, and $N^*$ contributions:
\be
\overline{\alpha}+\overline{\beta} &=&
(\overline{\alpha}+\overline{\beta})_{\rm loop}
+ \frac{8\alpha}9 \frac{m}{\md^2} \frac{\md^2+m^2}{\md^2-m^2} Q^2
+ \frac{2\alpha}3 \frac{m}{\ms^2} \frac{\ms^2+m^2}{\ms^2-m^2} Q_*{}^2
\, ,
\\
\overline{\beta} &=&
\overline{\beta}_{\rm loop}
+ \frac{8\alpha}9 \frac{Q^2}{\md-m}
- \frac{2\alpha}3 \frac{Q_*{}^2}{\ms+m}
\, .
\label{NS2}
\ee
Note that the $\Delta$ contribution gives $E2/M1\sim -0.3$ 
while empirically it is $-0.015$, and the $N^*$ contribution gives
$E1/M2\sim -3$ while empirically it is between $-2$ and $-3$. 
Using that the (isospin averaged) experimental results~\cite{PDG} are
$\frac12 (15.8+14.2)\times 10^{-4}$~fm$^3$
and $\frac12 (6.0+2.1)\times 10^{-4}$~fm$^3$ 
and the loop contributions~\cite{Meissner} are  
$\frac12(8.8+5.4)\times 10^{-4}$~fm$^3$
and $-\frac12 (1.5+2.2)\times 10^{-4}$~fm$^3$
respectively, we may estimate the values of the
couplings $Q$ and $Q_*$, with some freedom
available from error bars. We 
choose\footnote{The value of $Q=2.75$ fixed purely from the $\Delta$
as quoted 
in Ref.~\cite{US2} leads to an overestimation of the Compton cross section
when added to the background, 
as seen in the right plot of Fig.~\ref{rappcomp}. 
This comes from a double counting indicative of a treatment without
the proper unitarization of the background and resonances.}
$Q=1.8/m$ and $Q_*=2.2/m$ for which the polarizabilities are
$\overline{\alpha}+\overline{\beta}=13\times10^{-4}$~fm$^3$
and $\overline{\beta}=3.6\times10^{-4}$~fm$^3$.

We note that $R_*$ and $\overline{R}_*$ do not contribute 
at $q^2=0$, and are mainly associated with the longitudinal part of the
electroproduction cross section for $q^2<0$. By analogy with the 
$\Delta$, they will be set to zero~\cite{US2}. We are therefore left 
to constrain $S_*$ which is readily done by fitting to the Compton 
amplitude. Given the $\pi N$ background, we find $S_*=-1.5/m^2$. 
Similarly $S=1.5/m^2$ for the $\Delta$ offers a better fit than 
the value $S=1.2/m^2$ quoted in Ref.~\cite{US2}. 
The radiative width from the resonant part of the $\Delta$ and $N^*$ are
$0.27$ MeV and $0.19$ MeV respectively, to which the contribution from
the $\pi N$ background should be added.

\begin{figure}
\begin{center}
\leavevmode
\hbox{
\hspace*{-.4cm}
\epsfxsize=3.25in
\epsffile{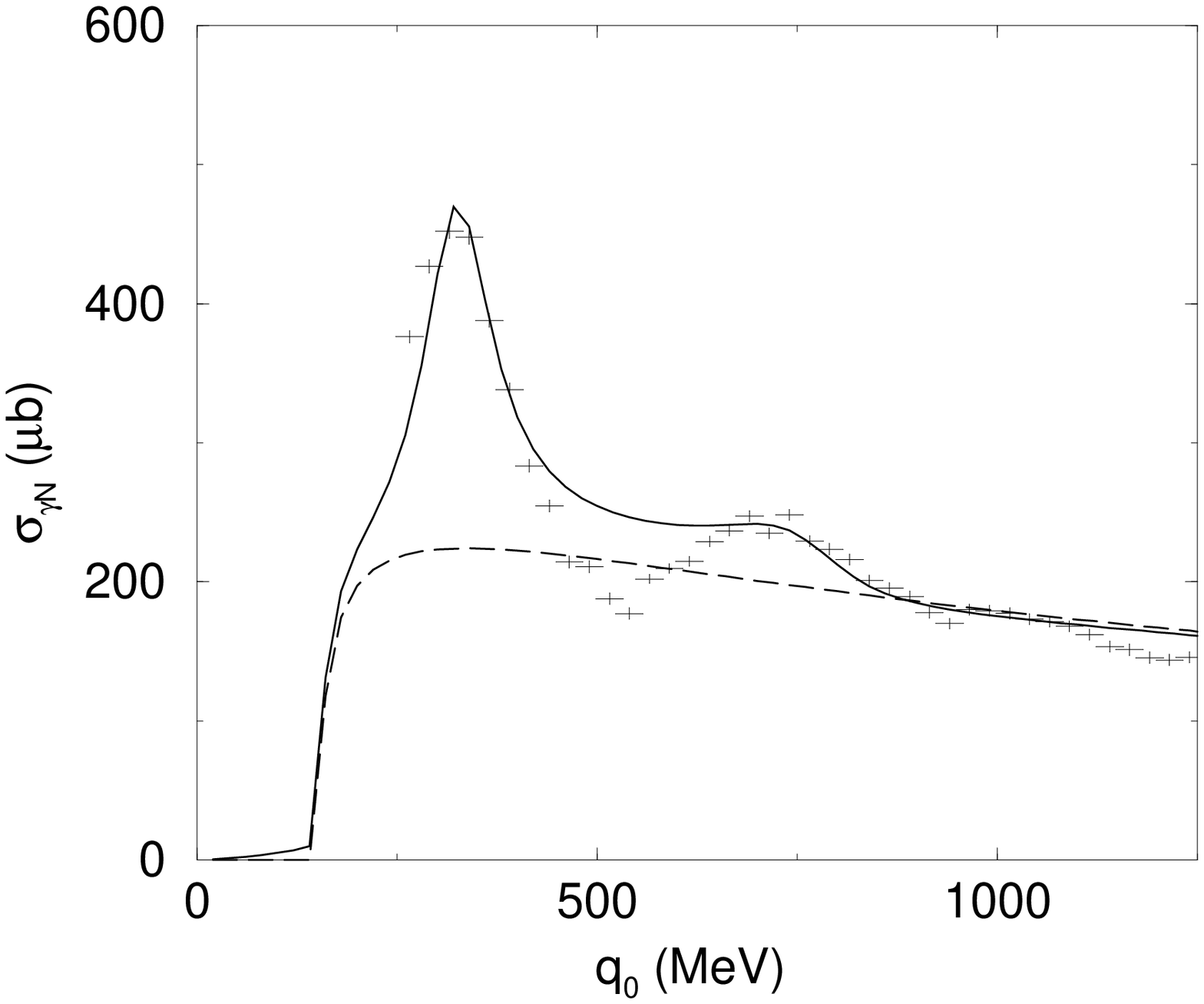}
\hspace{.1cm}
\epsfxsize=3.25in
\epsffile{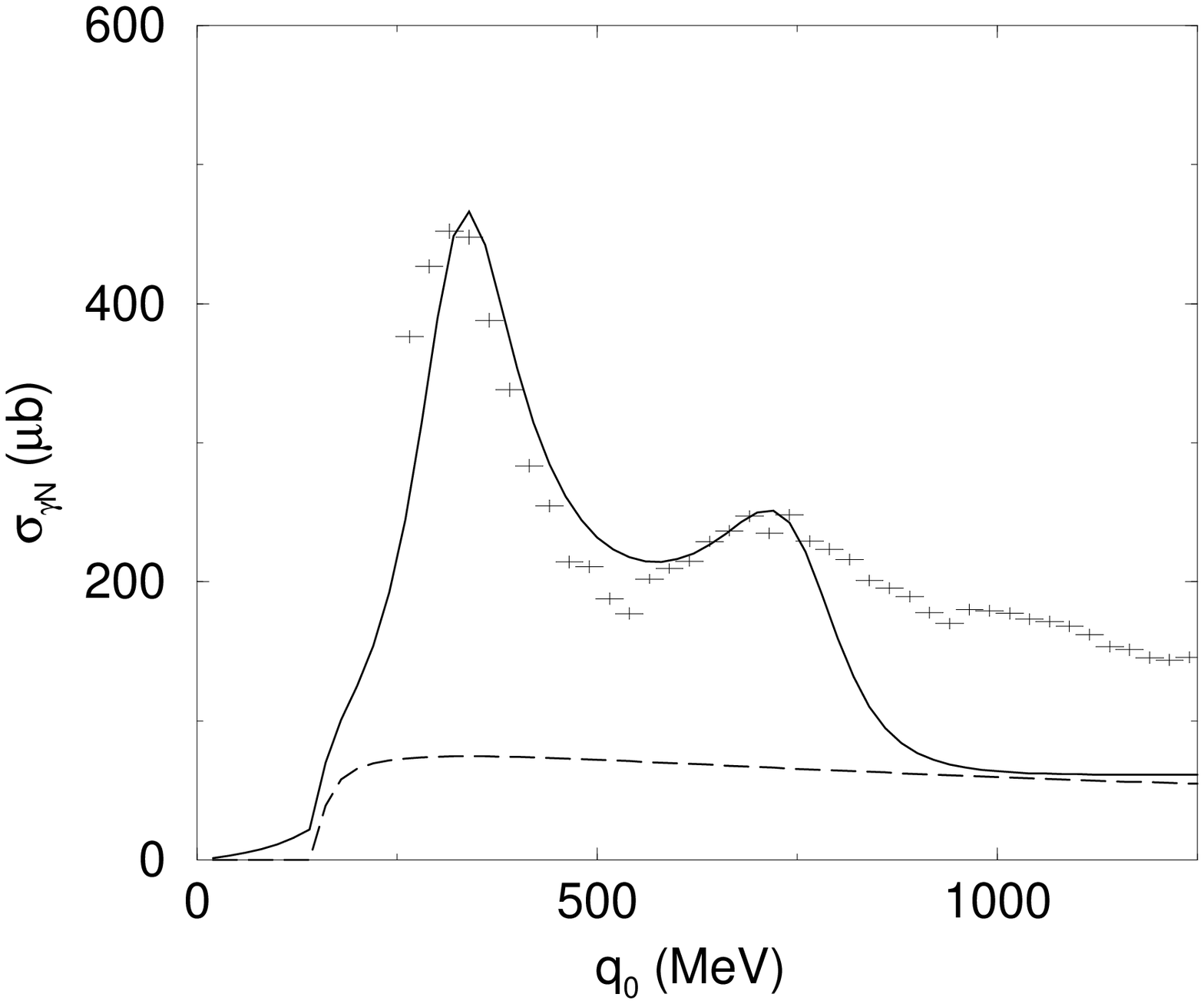}
}
\end{center}
\caption{\label{spec} 
Background contribution (dashed) and total contribution (solid)
to the Compton amplitude compared to data~\protect\cite{PDG}
(crosses). The left plot is for the full background and right plot is
for one-third the background.}
\end{figure}

The fit to the Compton amplitude resulting from this new set
of parameters and including the $\pi N$ and $\Delta$ background
is shown in the left plot of Fig.~\ref{spec}.  
The background contribution is
shown by the dashed line and the total contribution by the
solid line. The crosses
are the data~\cite{PDG}. The difference with Ref.~\cite{US2} is a
change in the  
$\Delta$ couplings (see above) to fit the data and the addition of
the $N^*$ to account for the 20\% discrepancy. The background is
parameter free.

To assess the importance of the $\pi N$ background versus
the $N^*$ resonance in the electromagnetic emission rates, it is
useful to do the following exercise: 
we drop the background by a factor of three and change the
$\Delta$ and $N^*$ parameters to recover the Compton amplitude, at
least through the $N^*$ region. 
This leads to the right plot of Fig.~\ref{spec}, again with
the background contribution shown by the dashed lines and the
total contribution by the solid line. The couplings are still fit to the
experimental polarizabilities in this case giving: 
$(Q,Q_*)=(1.8,4.3)/m$ and $(S,S_*)=(-2.6,-3.0)/m^2$. The
$E/M$-ratios are therefore the same as before.
This smaller background is consistent with the one used in
Ref.~\cite{RAPP1}.
This is clearly different from our approach in which the background is
first constrained by broken chiral symmetry.

\begin{figure}
\begin{center}
\leavevmode
\epsfxsize=4.25in
\epsffile{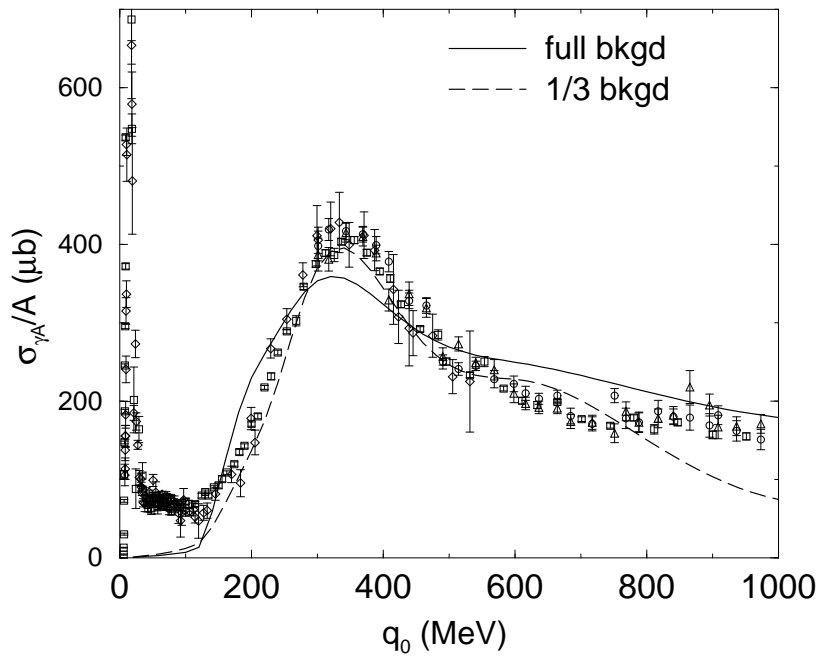}
\end{center}
\caption{\label{gammaA} Compton amplitude on nuclei comparing the two
background choices to data~\protect\cite{NUCLEI}.}
\end{figure}

The Compton amplitude on nuclei is also measurable~\cite{NUCLEI}. 
To leading order in the nucleon density it can be readily assessed
using the Compton amplitude on a single nucleon smeared by Fermi
motion (ignoring Pauli blocking),
\be
\frac{\sigma_{\gamma A}}{A} = \int\! \frac{d^3p}
{4\pi p_F^3/3}\; \theta(p_F-|p|)\;
\sigma_{\gamma N}(s) 
\label{NS4}
\ee
with $q^2=0$ and $s=m^2+2q_0 (E_p-|p| \cos\theta_p)$ and $p_F=265$~MeV.
The results from Eq.~(\ref{NS4})
are shown in Fig.~\ref{gammaA} versus the data\cite{NUCLEI}. The solid
line follows from the left plot of Fig.~\ref{spec} while the dashed
line from the right plot of Fig.~\ref{spec}. 
Since we do not account for particle-hole excitations
below the $\pi N$ threshold, we have not reproduced the large increase
near zero $q_0$.  This has no consequence for the dilepton rates to be 
discussed below.  
The discrepancy of the
dashed curve at high $q_0$ follows from neglecting the higher
resonances in Fig.~\ref{spec}.
Fermi motion smears the
contribution of the $\Delta$ and $N^* (1520)$ resonances more than the
background. 
The results for the smaller background (dashed line) is in better
agreement with the data, although both predictions are within one
standard deviation of the data.

\begin{figure}
\begin{center}
\leavevmode
\hbox{
\epsfxsize=3.25in
\epsffile{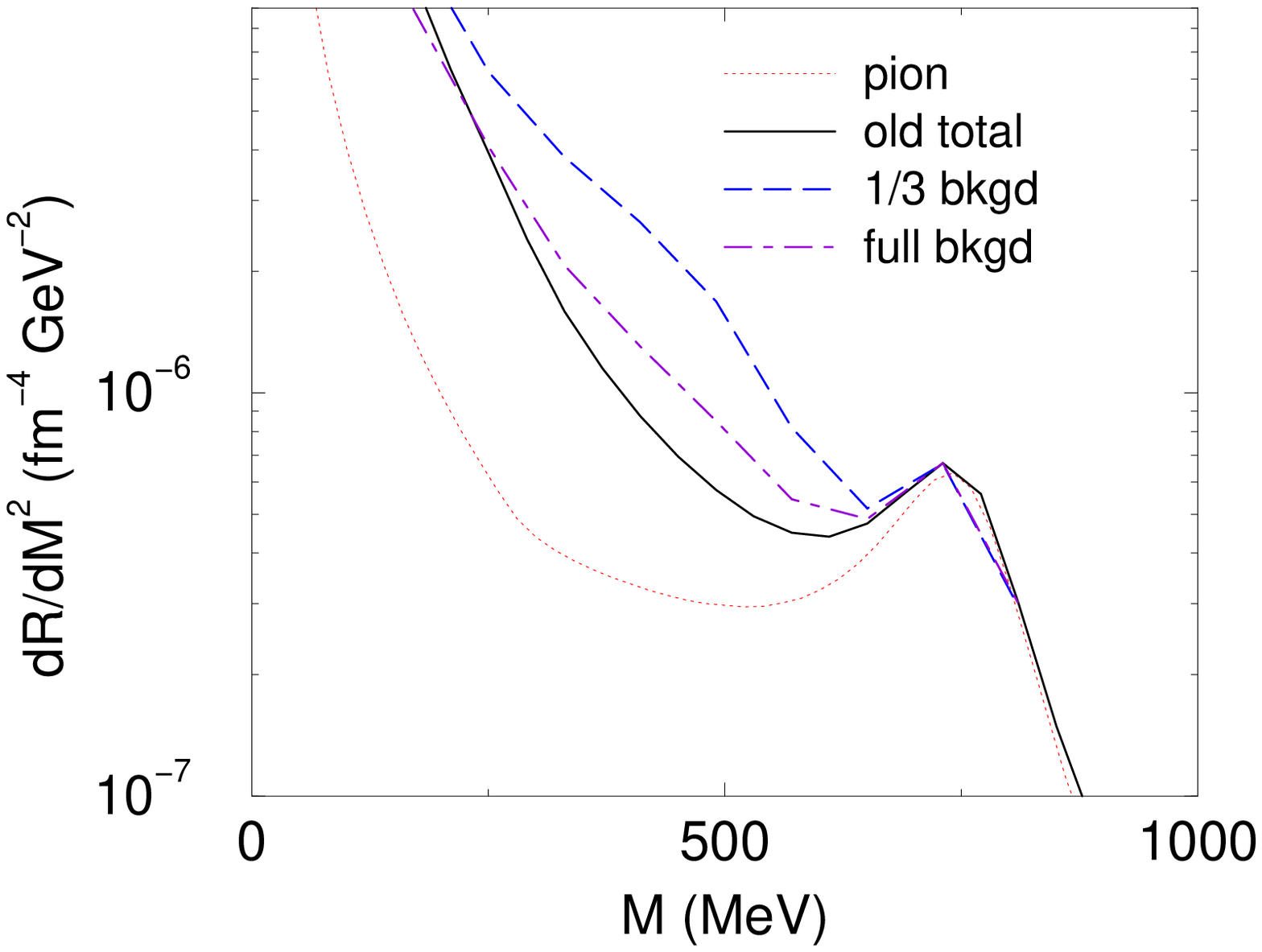}
\epsfxsize=3.25in
\epsffile{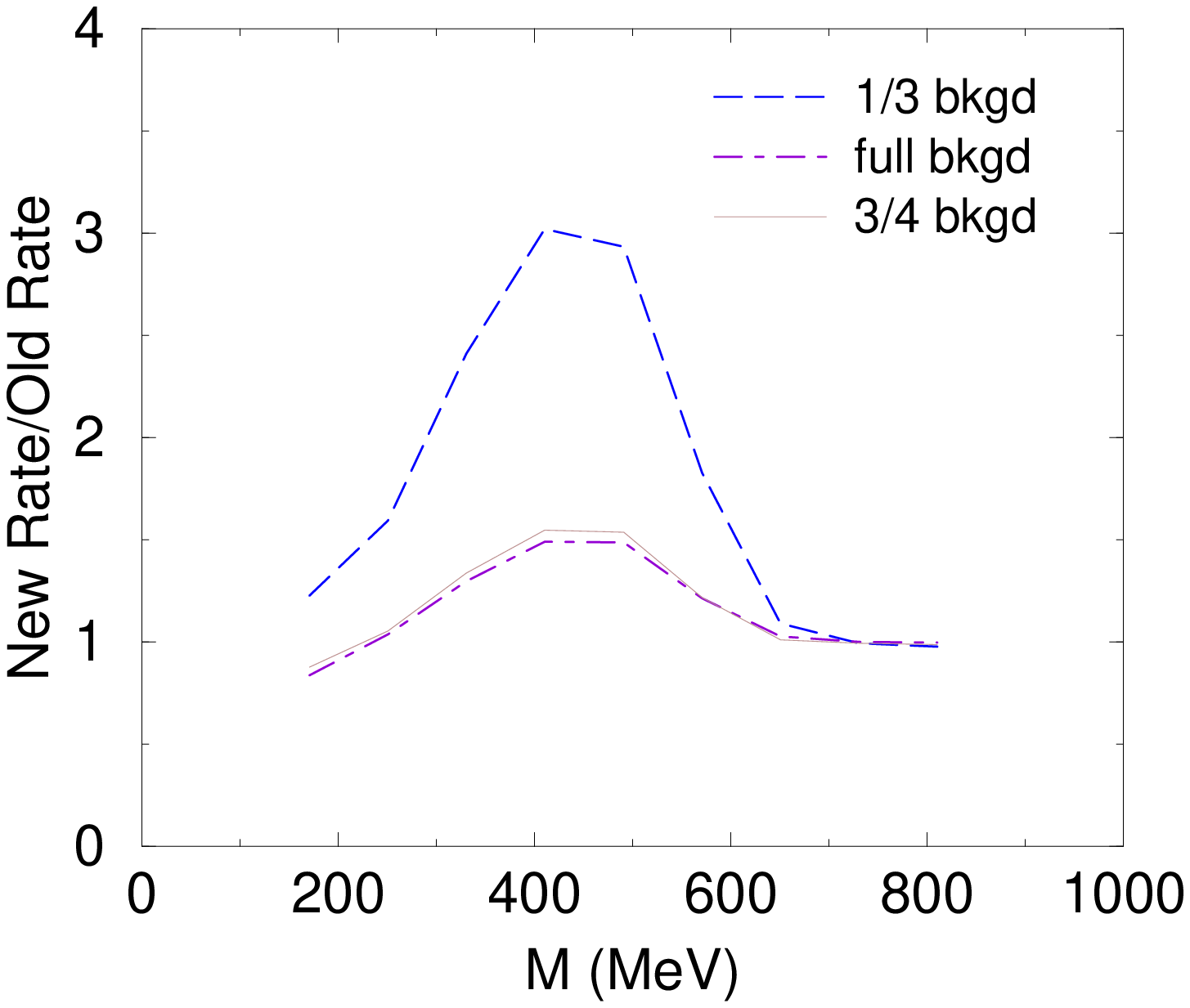}
}
\end{center}
\caption{\label{dilep}
The left shows the dilepton rate for $T=150$~MeV and $\mu=540$~MeV
comparing our old result with that of the one-third
background and full background.
The right shows the ratio of the new results to the old result.}
\end{figure}

The nucleon contribution to the Compton amplitude as shown in 
Fig.~\ref{spec} can be used in Eq.~(\ref{6}) in conjunction with the
pion contribution as discussed in Ref.~\cite{US2}. 
Using the identity ($M=\sqrt{q^2}$)
\ben
\frac{d{\bf R}}{d^4q} = \frac{2}{\pi} \frac{d{\bf R}}{dM^2 dy
dq^2_\perp}\, ,
\een
and integrating over rapidity $y$ and transverse momentum $q^2_\perp$,
the dilepton rates are shown in the left plot of 
Fig.~\ref{dilep} for $T=150$~MeV and $\mu=540$~MeV
(corresponding to a nucleon density of $\rho_0$).
The purely pionic contribution is shown by the dotted line,
the dot-dashed line is the result corresponding to the left plot of
Fig.~\ref{spec}, while the dashed line is the result corresponding to
the right plot of Fig.~\ref{spec}. The result quoted in
Ref.~\cite{US2} is shown 
by the solid line. As expected, the changes between the solid
and dot-dashed lines are small and within expectations. More
importantly, the dashed rate, following from the smaller background
contribution, is substantially larger due to the enhancement
caused by the additional strength of the $N^*$ resonance\footnote{
There is also an additional strength from the $\Delta$ resonance, but
it is significantly less.}.

To make further comparison between the dilepton rates from
different backgrounds,
the right plot of Fig.~\ref{dilep} shows the ratios of the new rates
to the old rate of Ref.~\cite{US2}. The dot-dashed line is for
the rate associated with the left plot of 
Fig.~\ref{spec} while the dashed line is for the
rate associated with the right plot of Fig.~\ref{spec}. 
It shows that in the
intermediate mass range, the enhancement of the rates obtained from a
smaller background \cite{RAPP1}
can be two times more than those from a 
larger\footnote{We have checked that an adjustment of the background 
to coincide with Ref.~\cite{RAPP}, 
which is about 75\% of that used in Ref.~\cite{US2}, 
yields a ratio
comparable to the full background as shown in the right plot of 
Fig.~\ref{dilep}.}
background \cite{US2}.
The photon emission rates show similar behavior to the dilepton rates,
with the smaller background giving twice as many photons
as the larger one.

Medium modifications as discussed in Ref.~\cite{RAPP,RAPP1} (and by
others~\cite{GERRY}) in addition appear to  
deplete the dilepton rates in the $\rho$ region, which is
qualitatively different from our rates with any background. 
The $N^*$ contribution is fully within the theoretical uncertainties for
our old rates as long as the full background is used.
This implies that in the low mass range, the rates are well accounted for 
by merely a dilute gas. The discrepancy between our rates~\cite{US2}
and those in Ref.~\cite{RAPP,RAPP1} around the $\rho$ is real and may be
indicative of some coherence or lack thereof. This point can be sorted out
through higher resolution measurements as we suggested in
Ref.~\cite{US1,US2}.

The electromagnetic emission rates in a baryon-free gas
are well established theoretically~\cite{GALE,US1}. In a
baryon-rich gas, the discrepancy between the rates reported
in Ref.~\cite{US2} and those in Ref.~\cite{RAPP1} is mainly due to the
choice of the $\pi N$ background. The
Compton amplitude on the nucleon and nuclei can both accommodate
either a 
small~\cite{RAPP1} or large~\cite{US2} $\pi N$ background.
A calculation with a 
small background requires more strength in the resonances, leading
to dilepton and photon rates that are about two times larger than from
a large 
background at $T\sim m_{\pi}$ and $\rho_N\sim \rho_0$.

Perturbative unitarity and broken chiral symmetry uniquely specify
the larger $\pi N$ background. A one-loop expression for this quantity was 
used here.  Adding more loops can alter and possibly deplete the background 
above the two-pion threshold.  However, these corrections are chirally
suppressed as evidenced by the fact that the one-loop calculation is
a good approximation to the background above the resonances. 
A systematic understanding of how the higher order contributions enter,
like through pion photon-production and pion knock-out, is still important.
This can be done reliably using the framework proposed in 
Ref.~\cite{MASTER}, where due care is paid to unitarity and broken chiral 
symmetry. As always, it is important to constrain the dynamics of relativistic
heavy ion collisions by broken chiral symmetry so that novel effects can be 
reliably assessed.

\vglue 0.6cm
{\bf \noindent  Acknowledgements \hfil}
\vglue 0.4cm
We thank our colleagues at both institutions for discussions,
in particular R.~Rapp for sharing some of his new calculations with us
as well as the data compilation from Fig.~\ref{gammaA}.
We also thank P.~Huovinen for providing us with the
data from Ref.~\cite{RAPP}, used in their comparative
analysis \cite{PRA}. We were also informed by R.~Rapp that conclusions
similar to ours were being reached~\cite{ONGOING}. 
This work was supported in part by the US DOE grant DE-FG02-88ER40388
and by the National Science Foundation under Grants No.~PHY-9511923
and PHY-9258270.

\end{document}